\documentclass[11pt,a4paper]{article}

\usepackage[margin=2.5cm]{geometry}
\usepackage[T1]{fontenc}
\usepackage[utf8]{inputenc}
\usepackage{lmodern}

\usepackage{cite}
\usepackage{amsmath,amssymb,amsfonts}
\usepackage{graphicx}
\usepackage{textcomp}
\usepackage{xcolor}
\usepackage{booktabs}
\usepackage{multirow}
\usepackage{url}
\usepackage{listings}
\usepackage{tikz}
\usepackage[most]{tcolorbox}
\usepackage{caption}
\usepackage{authblk}

\usepackage[hidelinks]{hyperref}
\usepackage{cleveref}

\usepackage{glossaries}
\glsdisablehyper 
\newacronym{LLM}{LLM}{Large Language Model}
\newacronym{ML}{ML}{Machine Learning}

\definecolor{ChatBeige}{HTML}{F7F0E6}
\definecolor{ChatBorder}{HTML}{D8CDBD}
\definecolor{ChatIcon}{HTML}{6F6A63}
\definecolor{DiffRed}{HTML}{B00020}
\definecolor{DiffGreen}{HTML}{006D32}

\newtcolorbox{chatpromptbox}{
  enhanced,
  colback=ChatBeige,
  colframe=ChatBorder,
  boxrule=0.3pt,
  arc=4pt,
  outer arc=4pt,
  width=\columnwidth,
  left=4pt,
  right=4pt,
  top=4pt,
  bottom=4pt,
  boxsep=0pt,
  before skip=0pt,
  after skip=8pt
}

\newcommand{\promptvar}[1]{\texttt{\textdollar #1}}
\newcommand{\diffminus}[1]{\textcolor{DiffRed}{\texttt{-#1}}}
\newcommand{\diffplus}[1]{\textcolor{DiffGreen}{\texttt{+#1}}}

\lstdefinestyle{prompt}{basicstyle=\scriptsize\ttfamily,breaklines=true,
  columns=fullflexible,frame=single,framesep=4pt,keepspaces=true,
  showstringspaces=false,aboveskip=4pt,belowskip=4pt}

\providecommand{\keywords}[1]{%
  \vspace{0.6em}\noindent\textbf{Index Terms---}\,#1\par\vspace{0.6em}}

\title{\bfseries Labelling Bug-Fixing Commits with Local\\
Open-Weight Language Models}

\author[1]{Philip König\thanks{Corresponding author: \texttt{philip.koenig@univie.ac.at}}}
\author[2]{Georg Goldenits}
\author[3]{Caroline König}
\author[2]{Sebastian Raubitzek}
\author[2]{Fabian Obermann}
\author[2]{Dennis Toth}
\author[1]{David Schmidt}
\author[1]{Edgar Weippl}
\author[2]{Kevin Mallinger}

\affil[1]{Faculty of Computer Science, University of Vienna, Vienna, Austria\\
\texttt{\{philip.koenig,caroline.koenig,d.schmidt,edgar.weippl\}@univie.ac.at}}
\affil[2]{SBA Research gGmbH, Floragasse 7/5.OG, 1040 Vienna, Austria\\
\texttt{\{ggoldenits,sraubitzek2,fobermann,dtoth,kmallinger\}@sba-research.org}}
\affil[3]{Christian Doppler Laboratory AsTra, University of Vienna, Vienna, Austria}

\date{}

\begin{document}
\maketitle

\begin{abstract}
Defect prediction depends on knowing which commits fix bugs, yet the labels that
encode this are produced by routes that each introduce noise. Reused benchmarks
carry documented data-quality problems, issue-tracker links are biased and the
underlying reports are frequently mistyped, and matching keywords in commit
messages is a coarse heuristic. This paper examines whether commits can be
labelled as bug fixes from their content alone, using open-weight language
models that run locally and therefore keep the process reproducible, inexpensive
at corpus scale, usable on proprietary code, and independent of any issue
tracker. Against datasets of manually validated and curated bug fixes spanning
Java, Python, and JavaScript, we compare a keyword baseline with a set of
open-weight models of varying size, prompting each with the commit message and
the code diff. On the manually validated corpus the keyword baseline recovers
fewer than half of the fixes, whereas the open-weight models recover the large
majority and outperform it repository by repository with statistical
significance, and larger models do not consistently outperform smaller ones. We
further show that evaluation corpora without negative examples cannot support a
precision-aware comparison of such classifiers. We release the labelling
pipeline together with a labelled, multi-language corpus produced by the
recommended configuration, as a reproducible silver-standard resource for
building current, project-specific datasets.
\end{abstract}

\keywords{bug-fix identification, defect prediction, data quality, label noise,
large language models, mining software repositories, reproducibility}

\section{Introduction}
\label{sec:intro}

Defect prediction directs testing and review effort towards the parts of a
codebase most likely to contain faults, and has become an established tool
in software engineering research and practice~\cite{rathore2019study, LiJingZhu2018}. Every such prediction model is trained and evaluated on data that rests on one prior decision, namely which commits in a project's history fixes bugs. That decision yields the bug-fix label, and it is the first link in a chain the rest of the pipeline inherits. A bug-fixing commit is the starting point from which the SZZ (\'Sliwerski--Zimmermann--Zeller) family of algorithms traces the change that introduced a defect~\cite{Sliwerski2005ChangesIO, kim2006automatic}, and the resulting bug-introducing labels are what train the models. Noise in the bug-fix label, therefore, does not stay local. It propagates into the inducing labels and into every model built on them. The labels, rather than the models, are the binding constraint on this line of work.

In current practice, the bug-fix label is produced by one of three routes, and
each introduces noise. The first is to reuse an established benchmark. The
widely used collections are old, as the NASA and PROMISE-era data sets date to
the 2000s~\cite{Menzies2007Promise}, and their quality has been questioned to
the point that careful cleaning is required before they are used~\cite{NASA_2013, 10.1007/s10462-022-10371-6}. They also predate much of
contemporary development practice, so assumptions calibrated on them need not
transfer to current projects. The second route links commits to bug reports in
an issue tracker. Raw linking carries two well-documented problems. Issue types
are misclassified, as a substantial share of reports filed as bugs do not
describe corrective work~\cite{Herzig2013Misclassification}, and the links are
incomplete, as many genuine fixes are never connected to any issue at
all~\cite{Bird2009FairBalanced}. The third route matches keywords such as
\emph{fix} or \emph{bug} in commit messages~\cite{Sliwerski2005ChangesIO}. It
is cheap to apply and therefore common as a first pass, but coarse, because
commit messages are short, inconsistent, and frequently silent about whether a
change repairs a defect.

This paper takes a different route. Instead of choosing among these noisy
sources, we label a commit from its own content, reading out the committed code
together with the commit message. The instrument is a set of \Glspl{LLM} run locally on fixed, openly available weights. This gives the
approach three properties the other routes lack. First, the labelling is
reproducible and auditable, as the weights are pinned and decoding is
deterministic. Re-running the process on the same commits returns the same
labels. Second, it is inexpensive at the scale of a full repository history and
never sends code outside the organization, so it can be applied to proprietary
projects that cannot be exposed to an external service. Third, it reads the
commit directly, which leaves it independent of the issue tracker and of the
linking step responsible for much of the noise described above. A team can
therefore label its own repositories and assemble a current data set from its
own history, rather than reusing an aging benchmark or inheriting the biases of
issue linking.

We do not claim that this produces perfect labels. No labelling method for this
task is an oracle, the one we study included, and rather than assume its error
away we measure it against validated data. The contribution is comparative and
transparent, namely labels that are more accurate than the cheap heuristic,
produced reproducibly, and reported together with their measured error, which we
treat throughout as a silver standard. The restriction to locally runnable
models is deliberate for the same reason. Hosted frontier models are not a
baseline here, not because they are weaker at the task, which we do not test, but because a durable reference corpus needs a labelling process that can be
regenerated and inspected years later. A hosted model is a moving target
that is versioned, deprecated, and updated without notice. Local open-weight models satisfy the reproducibility, cost, and data-governance constraints the use case imposes, whereas hosted models do not, and small open models run locally have recently been shown to be a viable alternative to hosted ones for related classification tasks~\cite{Goldenits2026SLM}.

We address the following three research questions:

\begin{itemize}
  \item[\textbf{RQ1}] How closely does keyword-based labeling reproduce
  manually validated bug-fix labels, across projects and languages?
  \item[\textbf{RQ2}] Can locally deployable open-weight language models label
  bug-fixing commits more accurately than the keyword baseline, and how does
  performance vary across models and prompts?
  \item[\textbf{RQ3}] How does the composition of the evaluation corpus, in
  particular the presence or absence of negative examples, affect the comparison
  of bug-fix classifiers?
\end{itemize}

To answer these questions we evaluate a keyword baseline and a set of
open-weight models spanning roughly an order of magnitude in parameter count
against data sets of manually validated and curated bug fixes in Java, Python,
and JavaScript, prompting each model with the commit message and the code diff.
The keyword baseline recovers fewer than half of the validated fixes, which
makes apparent how far the cheap default sits from the data it is meant to
approximate. The open-weight models recover the large majority of those fixes
and clearly outperform the baseline. A blocked, rank-based analysis confirms
that this separation holds repository by repository rather than only in
aggregate. Classification quality does not increase with parameter count across
the range we study, so the task does not require a model at the upper end of
what local hardware allows. Finally, the corpora that contain only positive examples rank the models very differently from the corpus that contains negatives, because without negatives a corpus cannot penalize a classifier that labels too many commits as fixes. This indicates that positives-only data cannot support a precision-aware comparison of classifiers for this task. We release the labeling pipeline together with a multi-language corpus of commits labeled by the recommended configuration, spanning a diverse set of open-source projects across the three languages, as a reproducible silver-standard resource for downstream use.
\section{Related Work}
\label{sec:related}

\subsection{Defect prediction and its ground truth}

Defect prediction has been studied for decades, and several surveys document
its methods, metrics, and predictive performance~\cite{rathore2019study,
LiJingZhu2018, DAmbros2012EMSE}. Models are built on features drawn from the
source code and from the development process, from size and complexity measures
to code churn and other change metrics~\cite{10.1145/1062455.1062514,
RahmanDevanbu2013Process}. Much of this effort concentrates on the features and
the choice of classifier, whereas the labels the models are trained on are
typically inherited from a heuristic and taken as given, a data-quality gap that
recent reviews have begun to draw attention to~\cite{10.1007/s10462-022-10371-6}.
Yet every supervised defect model depends on a prior decision about which
commits fix bugs, from which the defect labels are ultimately derived, and the
accuracy of the predictor is bounded by the accuracy of that initial labelling.
It is this step, rather than the modelling that follows, that the present work
addresses.

\subsection{Noise and quality in bug-fix and defect data}

The labels underlying defect data are known to be noisy, and the problem has
been documented from several directions. In a manual study of more than seven
thousand issue reports, a large fraction of those filed as bugs were found to
describe non-corrective work such as new features or refactorings, which biases
any model that treats the developer-assigned issue type as ground
truth~\cite{Herzig2013Misclassification}. Linking commits to issues introduces a
second problem, as the links are both biased and incomplete, and many genuine
fixes are never connected to an issue at all~\cite{Bird2009FairBalanced}. The
reused benchmarks on which much of the field has relied carry defects of their
own. The widely used NASA and PROMISE collections~\cite{Menzies2007Promise} have
documented data-quality problems serious enough that cleaned versions had to be
produced before they could be used reliably~\cite{NASA_2013,
shepperd_song_sun_mair_2018}. The downstream effect of such noise has been
measured directly, as injecting realistic label noise degrades defect models and
the recall of the resulting models is the dimension most
affected~\cite{Tantithamthavorn2015Mislabelling, Kim2011Noise}. A systematic
review of data-quality issues in fault prediction collects these and related
findings and concludes that label noise is a pervasive and under-addressed
threat~\cite{10.1007/s10462-022-10371-6}. These results motivate two choices in
this work. The first is to measure every method against manually validated
labels rather than against another heuristic. The second is to quantify the
error of the labels we produce against that reference and to release the
resulting corpus as a silver standard rather than presenting it as ground truth.

\subsection{Identifying bug-fixing commits}

In practice, bug-fixing commits are identified by one of a few routes. A
common and inexpensive one matches keywords such as \emph{fix} in the commit
message. The approach dates back to early work on identifying
fixes~\cite{Sliwerski2005ChangesIO} and remains in use today~\cite{Koenig2025BoostClassifier}. 
A second route links commits to typed
issues in a tracker, inheriting the bias and incompleteness noted
above~\cite{Bird2009FairBalanced}. These labels rarely remain an end in
themselves, as they seed the SZZ family of algorithms, which trace each fix back to the change suspected of introducing the defect~\cite{Sliwerski2005ChangesIO,
kim2006automatic}. This has been reimplemented and refined many times
since~\cite{Borg2019SZZUnleashed}, but its reliability depends heavily on the
quality of the fix set it starts from~\cite{RodriguezPerez2018SZZ}. The precision of the initial fix labelling therefore propagates into every bug-introducing label SZZ produces, which is one reason a labelling method intended to seed SZZ is best operated at a high-precision point.

Evaluating fix identification requires datasets of known bugs. Several recent
collections provide reproducible, manually checked bugs for individual
languages, namely BugsInPy~\cite{Widyasari2020BugsInPy} and
PyBugHive~\cite{Antal2024PyBugHive} for Python and BugsJS~\cite{Gyimesi2019BugsJS}
for JavaScript, but each contains only confirmed bug fixes and no negative
examples. The SmartSHARK dataset~\cite{Trautsch2020Ecosystem} is an exception, as
it provides both bug-fixing and non-bug-fixing commits for a set of Java
projects, with issue types and commit-issue links that have been manually
validated~\cite{Herbold2020IssueType}. This validation is what makes SmartSHARK
usable as a reference for precision, although its negative class is derived from
issue types rather than from inspecting each commit, a point we return to when
interpreting precision in \Cref{sec:results}.

More recently, language models have been applied to tasks adjacent to fix
labelling. One study uses \Glspl{LLM} to detect tangled commits at the
method level, deciding from the commit message and the method-level diff whether
a change is bug-related, with strong results from few-shot and chain-of-thought
prompting across a mix of proprietary and open models~\cite{Opu2025Tangled}.
Another applies a language model to the bug-introducing side, improving SZZ by
assessing blame candidates with additional context~\cite{Tang2025LLM4SZZ}. Both
support the premise of this paper, that a model reading code and messages judges
the nature of a change more reliably than a keyword does, but both address a
different target, method-level untangling in the first case and the inducing
commit in the second, rather than labelling whole commits as fixes.

Against this background, the contribution of this paper is a commit-level labelling method that runs locally and reproducibly on open-weight models and a released, multi-language corpus produced with it, addressing a gap that the heuristic, issue-tracker, and recent language-model approaches each leave open.
\section{Methodology}
\label{sec:methodology}

This section describes how the labeling method is evaluated. We compare a set
of locally deployable \glspl{LLM} with a keyword baseline that applies term
stemming to commit messages, across four datasets of human-verified bug fixes,
and assess the differences between them with a rank-based, blocked statistical
analysis. Each method receives a commit, its message together with its code
diff, and returns a binary label stating whether the commit is a bug fix.

\subsection{Labelling pipeline}
\label{sec:pipeline}

The pipeline takes a repository and assigns every commit a binary label
indicating whether it is a bug fix. It clones the repository and walks the
commit history. For each commit it extracts the commit message and the diff,
where the diff is limited to at most ten changed files and at most fifty lines
per file. A marker is appended whenever a diff is truncated, so the model is
aware when it sees a partial change. This bounds the size of each prompt, at the
cost of truncating unusually large commits, a limitation discussed in
\Cref{sec:threats}. The message and the bounded diff are placed into a fixed
prompt (\Cref{sec:prompts}) and sent to the model, which returns its decision in
a constrained format that is parsed into the binary label. Inference is
deterministic, with temperature $0$, top-$p$ $0$, and top-$k$ $1$, so the model
returns the same label for the same commit on every run. This is what allows a
labeled corpus to be regenerated exactly and audited, rather than only
reproduced in distribution. Labels are collected per repository. For the
evaluation they are joined with the reference labels of the datasets in
\Cref{sec:datasets}, and for the released corpus they are themselves the output.

\subsection{Model selection} \label{sec:models}

We evaluate seven locally deployable \Glspl{LLM} together with a \texttt{stemming} keyword baseline. The baseline applies an English-language stemmer to each commit message and labels the commit as a fix when a stemmed token matches one of the trigger terms \emph{fix}, \emph{bug}, \emph{fixup}, \emph{fail}, or \emph{correct}, following the keyword-based approach used in prior repository-mining work~\cite{Koenig2025BoostClassifier}. The \Glspl{LLM} were selected along two deliberate axes, parameter scale and code specialisation. All models were obtained from the Ollama model library\footnote{\url{https://ollama.com/library}} and downloaded on June 21st 2026.

\paragraph{Scale}
To assess how bug-fix classification performance varies with model capacity, we
fix the parameter range at roughly four points, namely a $\sim\!10$\,B class
(\texttt{codegemma:7b}\footnote{\url{https://ollama.com/library/codegemma}},
\texttt{gemma4:12b}\footnote{\url{https://ollama.com/library/gemma4}}), a
$\sim\!20$\,B class
(\texttt{codestral:22b}\footnote{\url{https://ollama.com/library/codestral}},
\texttt{mistral-small3.2:24b}\footnote{\url{https://ollama.com/library/mistral-small3.2}}),
a $\sim\!30$\,B class
(\texttt{qwen3-coder:30b}\footnote{\url{https://ollama.com/library/qwen3-coder}},
\texttt{qwen3.6:35b}\footnote{\url{https://ollama.com/library/qwen3.6}}), and one
substantially larger model that remains runnable on local hardware
(\texttt{qwen3.5:122b}\footnote{\url{https://ollama.com/library/qwen3.5}}).

\paragraph{Code specialisation}
Each general-purpose model in our set has a recently released code-specialised counterpart, and vice versa. Because the task operates on commits, that is, on code and code-adjacent text, we pair general and code-specialised models of comparable scale, for example \texttt{gemma4}/\texttt{codegemma}, \texttt{mistral-small3.2}/\texttt{codestral}, and \texttt{qwen3}/\texttt{qwen3-coder}.

\paragraph{Local deployability}
All selected models run locally without a hosted API, and the upper end of the scale axis (\texttt{qwen3.5:122b}) is bounded by what remains feasible on local hardware.

\subsection{Prompt design}
\label{sec:prompts}

Each model is evaluated under two fixed prompts, identical across all models, denoted P1 and P2. Holding the prompt constant across models ensures that observed differences are attributable to the models rather than to model-specific prompt tuning.

\paragraph{Prompt P1}
P1 was obtained through an iterative refinement process, which evaluated candidate prompts, their outputs inspected on a development subset, and the wording adjusted until the classification behaviour was satisfactory. The resulting prompt is shown in \Cref{fig:p1}.

\begin{figure}[h]
\centering
\begin{chatpromptbox}
{\scriptsize
You are a senior software engineer and your job is to review this commit and decide whether it is a bugfix, or not.  Bugfixes are commits whose primary purpose is to correct incorrect behavior, for example: fixing a crash, wrong output, logic error, regression, or edge-case failure. \\

The following examples are \textbf{NOT} bugfixes: new features, refactoring, performance optimization, documentation, formatting/style, dependency bumps. If a change does several things, classify by its primary purpose. \\

Here is the commit to judge: The commit modified a total of \texttt{\$totalFiles} files and has the following commit message: ``\texttt{\$commitMessage}''. \\

Here is the git diff output of this commit, to decide if this is a bugfix or not: \texttt{\$fileLines}

\texttt{\$formatInstruction}
}
\end{chatpromptbox}
\caption{The zero-shot prompt P1, giving the model a role, a one-line definition
of a bug fix, a short list of change types that are not fixes, and the commit
under test. Variables in typewriter font are substituted per commit.}
\label{fig:p1}
\end{figure}

\paragraph{Prompt P2}
P2 keeps the task of P1 but differs from it in several respects at once. It
states the definition of a bug fix more fully, adds an extended list of changes
that are not fixes, instructs the model to judge from the diff rather than the
wording of the message, and includes six pre-labeled input examples. The six examples are deliberately weighted toward non-fixes, two positive and four negative rather than balanced, because the harder and more frequent error for both the baseline
and the models is to label a non-fix as a fix when its message contains the word
\emph{fix}. The negative examples therefore cover the cases that most often
trigger that error, a version bump, a test-only change, a refactoring, and a
build change. Because P2 changes more than one thing relative to P1, the two are
not a single-variable manipulation, and we do not attribute any difference
between them to a particular change. The comparison serves two purposes instead.
It tests whether the ordering of the models is stable across two substantially
different prompts, and it provides two operating points, since the additional
instruction and examples in P2 move the models toward precision and away from
recall. The released corpus (\Cref{sec:datasets}) is labelled with P2, because
labels intended to seed SZZ-based analyses benefit from the precision-leaning
point, where a commit labelled a fix is more likely to be one and fewer spurious
fixes propagate into the downstream inducing labels~\cite{RodriguezPerez2018SZZ}.
We treat this as a property of the intended use, not as a claim that precision
is preferable in general. The resulting prompt, with two of its six examples, is
shown in \Cref{fig:p2}.

\begin{figure}[h]
\centering
\begin{chatpromptbox}
{\scriptsize
\noindent You are a senior software engineer reviewing one commit. Decide whether the commit is a bug fix.

\vspace{3pt}
\noindent A bug fix is a commit whose primary purpose is to repair a defect, meaning behaviour that was supposed to work but did not, such as a crash, a wrong result, a logic error, a regression, or an edge-case failure, among other defects. These examples are illustrative and not exhaustive. The change repairs existing intended behaviour rather than adding or changing intended behaviour.

\vspace{3pt}
\noindent The following are \textbf{NOT} bug fixes, even when the message contains the word ``fix'': new features, refactoring, performance work, documentation or comment edits, formatting or style, renaming, configuration changes, build or CI changes, dependency updates, version-number bumps, commits that only add or change tests without touching production code, reverts, and merge commits. This list is likewise illustrative and not exhaustive.

\vspace{3pt}
\noindent Judge from what the diff does, not from the wording of the message. The word ``fix'' is not evidence on its own. When the message and the diff disagree, trust the diff. If a commit does several things, call it a bug fix only when repairing a defect is its primary purpose.

\vspace{4pt}
\noindent\textbf{Examples follow. Each shows a message, a short diff, and the answer.}

\vspace{3pt}
\noindent\textbf{Message:} ``Fix crash when config file is missing''

\noindent\textbf{Diff:}

\noindent\texttt{- ConfigLoader.java: +4/-1}

\noindent\hspace*{1em}\diffminus{\hspace{0.5em}        return Files.readAllLines(p);}

\noindent\hspace*{1em}\diffplus{\hspace{0.5em}        if (!Files.exists(p)) \{}

\noindent\hspace*{1em}\diffplus{\hspace{0.5em}            return Collections.emptyList();}

\noindent\hspace*{1em}\diffplus{\hspace{0.5em}        \}}

\noindent\hspace*{1em}\diffplus{\hspace{0.5em}        return Files.readAllLines(p);}

\noindent\textbf{Answer:} \texttt{\{"bugfix": true\}}

\vspace{3pt}
\noindent\textbf{Message:} ``fix flaky timeout in retry test''

\noindent\textbf{Diff:}

\noindent\texttt{- RetryServiceTest.java: +1/-1}

\noindent\hspace*{1em}\diffminus{\hspace{0.5em}        client.setTimeout(100);}

\noindent\hspace*{1em}\diffplus{\hspace{0.5em}        client.setTimeout(2000);}

\noindent\textbf{Answer:} \texttt{\{"bugfix": false\}}

\vspace{4pt}
\noindent Here is the commit to judge: The commit modified a total of \promptvar{totalFiles} files and has the following commit message: ``\promptvar{commitMessage}''.

\vspace{3pt}
\noindent\promptvar{fileLines}

\vspace{3pt}
\noindent\promptvar{formatInstruction}
}
\end{chatpromptbox}
\caption{The few-shot prompt P2, showing the instruction block and two of its six pre-labeled input examples. The remaining four are provided with the artifact.}
\label{fig:p2}
\end{figure}

\subsection{Datasets}
\label{sec:datasets}

We use two collections of data, one to evaluate the labeling methods and one
that the method produces and we release.

\paragraph{Evaluation data}
The methods are evaluated against four datasets of human-verified bug fixes.
\textsc{bugsinpy}~\cite{Widyasari2020BugsInPy} and
\textsc{pybughive}~\cite{Antal2024PyBugHive} for Python and
\textsc{bugsjs}~\cite{Gyimesi2019BugsJS} for JavaScript consist only of
confirmed bug-fixing commits and contain no negative examples, so on these
datasets recall can be measured but precision cannot. \textsc{smartshark} is the
exception, as it provides both bug-fixing and non-bug-fixing commits for a set
of Java projects~\cite{Trautsch2020Ecosystem}, and it is therefore the only
dataset on which precision and the metrics derived from it are meaningful. Its
labels rest on manual validation of issue types and of the links between commits
and issues~\cite{Herbold2020IssueType}. Of the 39 \textsc{smartshark} projects
we use, 34 have manually validated issue types and the remaining five do not, a
distinction we draw on in \Cref{sec:results}. Because the positive labels derive
from validated bug-typed issues while the negative class is the complement, a
commit that repairs a defect without being linked to a bug issue is counted as a
negative, so the precision measured against \textsc{smartshark} is a
conservative lower bound rather than a true precision.

\paragraph{Released corpus}
Applying the pipeline with the recommended model, \texttt{mistral-small3.2:24b},
under prompt P2 yields the corpus we release. It spans 33 open-source projects
across Java, Python, and JavaScript and several application domains, including
security, healthcare, data processing, developer tooling, and web development,
and the projects range in size from a few dozen commits to tens of thousands, so
that very small projects are represented alongside large ones. \Cref{tab:corpus}
summarises its composition. The labels carry the measured error of the labeller
rather than human verification, which is why we describe the corpus as a silver
standard.

\begin{table}[t]
\centering
\caption{Composition of the released silver-standard corpus, summarised by
language. Labels are produced by \texttt{mistral-small3.2:24b} under prompt P2;
the full per-project list is provided with the artifact.}
\label{tab:corpus}
\begin{tabular}{lrrr}
\toprule
Language & Projects & Commits & Labelled fixes \\
\midrule
Java        & 22 & 302{,}095 & 69{,}856 \\
Python      &  8 &  40{,}065 &  3{,}880 \\
JavaScript  &  3 &  47{,}407 &  5{,}960 \\
\midrule
Total       & 33 & 389{,}567 & 79{,}696 \\
\bottomrule
\end{tabular}
\end{table}

\subsection{Evaluation and statistical analysis} \label{sec:stats}

For each model, prompt, and repository, we report accuracy, precision, recall, and F1-Score. Because only the \textsc{smartshark} corpus contains negatively labelled commits, precision and the metrics derived from it are meaningful only on that corpus (\Cref{sec:results}). We therefore conduct the formal model comparison on \textsc{smartshark}, treating its repositories as the units of evaluation.

To compare the eight classifiers across repositories, we follow the established protocol for comparing multiple classifiers across multiple datasets, as recommended by \cite{demsar2006Analysis}. The design is a complete block design, as every model is evaluated on the same set of repositories, so each repository is a block that influences all models simultaneously (some repositories are intrinsically harder than others). This structure, the number of models compared, and the non-normal, bounded nature of per-repository F1-Scores jointly motivate a non-parametric, rank-based, blocked analysis rather than parametric or unpaired alternatives.

\paragraph{Omnibus test}
We first apply the \emph{Friedman rank-sum test}, the non-parametric analogue of a repeated-measures ANOVA for blocked data. For each repository, it ranks the eight models by F1-Score and tests whether the models' average ranks differ more than would be expected by chance. Operating within-repository ranks neutralises the confound of differing repository difficulty, and, being rank-based, it makes no normality assumption, both of which are appropriate given that per-repository F1-Score is bounded in $[0,1]$, is skewed, and includes boundary values. The Friedman test provides a single omnibus decision, stating that pairwise comparisons are warranted only if it rejects the null of equal performance.

\paragraph{Post-hoc test}
When the omnibus test is significant, we apply the \emph{Nemenyi post-hoc test} to identify which pairs of models differ. Nemenyi is the counterpart to the Friedman test for all-pairwise comparisons, as it derives a single \emph{critical difference} (CD) in average ranks such that two models differ significantly if and only if their average ranks differ by more than the CD. Crucially, it incorporates the multiple-comparison correction by construction, which a naive battery of pairwise tests (e.g.\ repeated Wilcoxon signed-rank tests) does not. With eight models, there are 28 pairwise comparisons, and uncorrected testing would substantially inflate the family-wise error rate. Reporting a single CD also yields a coherent, transitive-where-possible grouping of models into statistically indistinguishable sets, rather than an unordered collection of pairwise verdicts.

We perform this analysis separately for P1 and P2 to assess the stability of conclusions about model ordering across prompts. One caveat of the chosen design is noted in that the analysis weights every repository equally, regardless of its commit count, consistent with the macro-averaging used throughout, but not a commit-volume-weighted statement.

For \textsc{smartshark} we report descriptive per-repository statistics in two
forms, on the 34 projects whose issue types are manually validated and on the
full project set. The validated subset provides the primary precision and recall
figures, since precision is only as reliable as the validation behind the
negative class, and the full set is reported alongside it to show how much the
unvalidated projects change the picture. The omnibus Friedman test and the
Nemenyi post-hoc analysis are computed over the full set of projects, as are the
per-repository maps.

\subsection{Technical Detail}
All experiments were run on two rented L40 GPUs available on runpod\footnote{\url{https://runpod.io/}}. The statistical analysis was performed in R version 3.6.1.
\section{Results and Discussion}
\label{sec:results}

We evaluate eight bug-fix commit classifiers, the seven \Glspl{LLM} and a \texttt{stemming} keyword baseline, across the four benchmark data corpora \textsc{bugsinpy}, \textsc{bugsjs}, \textsc{pybughive}, and \textsc{smartshark}. Each \Gls{LLM} is run on two prompts, denoted \textbf{P1} and \textbf{P2}, which are held constant across all models. This lets us separate model effects from prompt effects and assess the robustness of our conclusions to prompt design.

Due to the data structure, where only the \textsc{smartshark} dataset contains positive and negative labels, for the other datasets, a false positive is not representable and precision is pinned near one by construction. As a consequence, F1-Score on \textsc{bugsinpy}, \textsc{bugsjs}, and \textsc{pybughive} reduces to a monotone function of recall and cannot distinguish a discriminating classifier from one that labels every commit as a bug fix. We use this property as a lens for assessing benchmark validity and treat \textsc{smartshark} as the only corpus for which the full set of metrics is meaningful.

\subsection{Per-corpus performance and the benchmark-validity inversion} \label{sec:percorpus}

\Cref{tab:percorpus} reports mean F1-Scores per dataset (averaged over the repositories within each corpus) for both prompts. The result is a striking inversion that is \emph{robust to prompt choice}: the corpus ranking of the classifiers on the three bug-fix-only corpora is close to reversed on \textsc{smartshark}. The two classifiers that label almost every commit a fix, \texttt{codegemma:7b} and \texttt{codestral:22b}, which for that reason we set aside as degenerate, make the mechanism concrete, as they top all three bug-fix-only corpora yet fall to last on \textsc{smartshark}. The effect is general, however, because the bug-fix-only corpora cannot penalise false positives and therefore reward recall regardless of precision.

\begin{table}[htbp]
\centering
\caption{Mean F1-Scores per corpus (averaged over repositories) for both prompts. The corpus ranking on the bug-fix-only corpora is close to reversed on \textsc{smartshark} (the only corpus with negative labels). This inversion is robust across prompts.}
\label{tab:percorpus}
\setlength{\tabcolsep}{4pt}
\resizebox{\textwidth}{!}{%
\begin{tabular}{ll cccccccc}
\toprule
Prompt & Corpus & \texttt{mistral} & \texttt{qwen3.6:35b} & \texttt{qwen3.5:122b} & \texttt{qwen3.coder} & \texttt{gemma4} & \texttt{codestral} & \texttt{codegemma} & \texttt{stemming} \\
\midrule
\multirow{4}{*}{P1}
 & \textsc{bugsinpy}   & 0.883 & 0.929 & 0.950 & 0.938 & 0.957 & 0.967 & \textbf{0.999} & 0.829 \\
 & \textsc{bugsjs}     & 0.771 & 0.873 & 0.957 & 0.905 & 0.945 & 0.936 & \textbf{0.998} & 0.792 \\
 & \textsc{pybughive}  & 0.886 & 0.940 & 0.945 & 0.925 & 0.916 & 0.966 & \textbf{0.995} & 0.657 \\
 & \textsc{smartshark} & \textbf{0.630} & 0.617 & 0.602 & 0.596 & 0.584 & 0.503 & 0.396 & 0.380 \\
\midrule
\multirow{4}{*}{P2}
 & \textsc{bugsinpy}   & 0.848 & 0.837 & 0.911 & 0.840 & 0.861 & 0.934 & \textbf{0.998} & 0.829 \\
 & \textsc{bugsjs}     & 0.735 & 0.673 & 0.833 & 0.697 & 0.771 & 0.919 & \textbf{0.988} & 0.792 \\
 & \textsc{pybughive}  & 0.817 & 0.860 & 0.915 & 0.773 & 0.859 & 0.925 & \textbf{0.997} & 0.657 \\
 & \textsc{smartshark} & \textbf{0.632} & 0.619 & 0.625 & 0.613 & 0.618 & 0.528 & 0.412 & 0.385 \\
\bottomrule
\end{tabular}}
\end{table}

\subsection{Repository-level analysis on \textsc{smartshark}} \label{sec:smartshark}

Because \textsc{smartshark} is the only corpus on which precision, recall, and F1-Score are jointly meaningful, and because its repositories are large (ranging from roughly one hundred to several thousand commits), we treat it as the primary evaluation. \Cref{tab:smartshark} reports, for both prompts, the mean and standard deviation across these three metrics.

\begin{table}[htbp]
\centering
\caption{Per-repository F1, precision, and recall on the 34 manually validated
\textsc{smartshark} projects (mean $\pm$ SD across repositories) for both
prompts, sorted by P1 mean F1. P2 shifts the operating point toward precision,
and the precision-optimal model is prompt-dependent.}
\label{tab:smartshark}
\setlength{\tabcolsep}{3pt}
\renewcommand{\arraystretch}{1.15}
\resizebox{\textwidth}{!}{%
\begin{tabular}{l ccc ccc}
\toprule
& \multicolumn{3}{c}{Prompt P1} & \multicolumn{3}{c}{Prompt P2} \\
\cmidrule(lr){2-4}\cmidrule(lr){5-7}
Model & F1 & Prec. & Rec. & F1 & Prec. & Rec. \\
\midrule
\texttt{mistral-small3.2:24b} & $\mathbf{0.645}{\pm}0.093$ & $\mathbf{0.572}{\pm}0.107$ & $0.751{\pm}0.099$ & $\mathbf{0.644}{\pm}0.085$ & $0.588{\pm}0.105$ & $0.724{\pm}0.092$ \\
\texttt{qwen3.6:35b}          & $0.633{\pm}0.094$ & $0.545{\pm}0.117$ & $0.773{\pm}0.088$ & $0.633{\pm}0.085$ & $\mathbf{0.634}{\pm}0.105$ & $0.644{\pm}0.106$ \\
\texttt{qwen3.5:122b}         & $0.620{\pm}0.093$ & $0.517{\pm}0.117$ & $0.799{\pm}0.074$ & $0.641{\pm}0.084$ & $0.602{\pm}0.115$ & $0.703{\pm}0.095$ \\
\texttt{qwen3-coder:30b}      & $0.609{\pm}0.095$ & $0.493{\pm}0.109$ & $0.818{\pm}0.073$ & $0.629{\pm}0.090$ & $0.618{\pm}0.105$ & $0.650{\pm}0.105$ \\
\texttt{gemma4:12b}           & $0.600{\pm}0.096$ & $0.488{\pm}0.119$ & $0.808{\pm}0.075$ & $0.636{\pm}0.082$ & $0.606{\pm}0.109$ & $0.685{\pm}0.101$ \\
\texttt{codestral:22b}        & $0.519{\pm}0.112$ & $0.374{\pm}0.118$ & $0.904{\pm}0.062$ & $0.546{\pm}0.103$ & $0.405{\pm}0.117$ & $0.879{\pm}0.061$ \\
\texttt{codegemma:7b}         & $0.410{\pm}0.126$ & $0.268{\pm}0.115$ & $0.983{\pm}0.015$ & $0.429{\pm}0.122$ & $0.286{\pm}0.115$ & $0.953{\pm}0.029$ \\
\texttt{stemming}             & $0.399{\pm}0.114$ & $0.436{\pm}0.137$ & $0.394{\pm}0.143$ & $0.399{\pm}0.114$ & $0.436{\pm}0.137$ & $0.394{\pm}0.143$ \\
\bottomrule
\end{tabular}}
\end{table}

As set out in \Cref{sec:datasets}, five of the \textsc{smartshark} projects lack manually validated issue types, so their reference labels are themselves model-predicted rather than checked. We therefore report the descriptive statistics on the 34 validated projects, where precision rests on a validated negative class. Reintroducing the five unvalidated projects lowers every method's mean F1 and widens its spread, as \texttt{mistral-small3.2:24b} falls from 0.645 to 0.630 under P1 with its standard deviation rising from 0.09 to 0.11, and the baseline falls from 0.399 to 0.380. The degradation is concentrated in those five projects and is largest for \texttt{commons-rdf}, the same project that stands out as the visible anomaly in the per-repository maps below. This is the issue-type mistyping of \cite{Herzig2013Misclassification} entering directly through the unvalidated labels rather than through the predictors, and it is why we treat the validated 34 as the reference for precision. The ordering of the classifiers and their margin over the baseline are unchanged, and the omnibus test and the maps below use the full project set.

\paragraph{F1-Score}
Under both prompts the same group of \Glspl{LLM} tops the ranking, with \texttt{mistral-small3.2:24b} highest (0.645 under P1, 0.644 under P2). As the across-repository standard deviation ($\approx 0.09$) is large relative to the gaps between adjacent models, we test the ordering rather than reading it from the means directly. Treating each \textsc{smartshark} repository as a block, a Friedman rank-sum test over the eight models (P1 with $N=38$ repositories, one repository excluded for an incomplete row) strongly rejects the null of equal performance ($\chi^2(7)=222.7$, $p<2.2\times10^{-16}$). We follow it with a Nemenyi all-pairs post-hoc test, whose critical difference in average ranks is $\mathrm{CD}=1.703$, as shown in \Cref{tab:ranks}.

\begin{table}[htbp]
\centering
\caption{Average Friedman ranks over the \textsc{smartshark} repositories (lower is better) for both prompts. Models whose ranks differ by less than the Nemenyi critical difference ($\mathrm{CD}$) are not significantly different. Under P1 a top pair is separable, while under P2
the five leading models form a single indistinguishable group. P1: $N=38$, $\mathrm{CD}=1.703$. P2: $N=38$, $\mathrm{CD}=1.681$.}
\label{tab:ranks}
\begin{tabular}{lcc}
\toprule
Model & P1 rank & P2 rank \\
\midrule
\texttt{mistral-small3.2:24b} & \textbf{1.58} & \textbf{2.64} \\
\texttt{qwen3.6:35b}          & 2.26 & 3.15 \\
\texttt{qwen3.5:122b}         & 3.20 & 2.91 \\
\texttt{qwen3-coder:30b}      & 3.76 & 3.59 \\
\texttt{gemma4:12b}           & 4.30 & 3.27 \\
\texttt{codestral:22b}        & 6.11 & 5.59 \\
\texttt{codegemma:7b}         & 7.37 & 7.33 \\
\texttt{stemming}             & 7.42 & 7.51 \\
\bottomrule
\end{tabular}
\end{table}

The post-hoc analysis sharpens the picture beyond what the means alone permit. \texttt{mistral-small3.2:24b} attains the best average rank (1.58) and is statistically indistinguishable from \texttt{qwen3.6:35b} (2.26) and \texttt{qwen3.5:122b} (3.20), even though the latter sits at the margin of the critical difference. These form the top group. At the same time, \texttt{mistral-small3.2:24b} is \emph{significantly} better than \texttt{qwen3-coder:30b}, \texttt{gemma4:12b}, and all lower-ranked models ($p<0.003$ in each case), so it is not merely co-best but separable from the lower half of the language-model field. The keyword baseline and the two degenerate classifiers occupy the bottom three ranks (6.11, 7.37, 7.42), each significantly worse than every model in the top group and mutually indistinguishable. The \Glspl{LLM} thus significantly outperform the baseline.\footnote{Nemenyi intervals are not transitive: a model may be indistinguishable from a neighbour yet significantly better than a model the neighbour ties. We therefore report ``not significantly distinguishable from'' rather than equality throughout. The protocol follows Dem\v{s}ar~(2006).}

The same analysis under P2 ($N=38$ repositories) is likewise globally significant ($\chi^2(7)=184.86$, $p<2.2\times10^{-16}$, $\mathrm{CD}=1.681$), but yields a markedly different separation structure. Under P2 the five leading language models, which we refer to collectively as the leading cluster, are mutually indistinguishable, as their average ranks span only 2.64 to 3.59 (\texttt{mistral-small3.2:24b}, \texttt{qwen3.5:122b}, \texttt{qwen3.6:35b}, \texttt{gemma4:12b}, \texttt{qwen3-coder:30b}), a range below the critical difference, and every pairwise comparison among them is far from significant ($p>0.68$). The discriminating power that separated \texttt{mistral-small3.2:24b} from the lower language models under P1 therefore disappears under P2. So the prompt that shifts the operating point toward precision (\Cref{sec:smartshark}) also \emph{compresses} the leading models into a single statistically indistinguishable group. The bottom three ranks are again occupied by \texttt{codestral:22b} (5.59), \texttt{codegemma:7b} (7.33), and \texttt{stemming} (7.51), each significantly worse than all five leaders, so the language models significantly outperform the baseline under both prompts as shown in \Cref{tab:ranks}.

\paragraph{Precision and recall}
The precision/recall decomposition exposes a clear prompt effect. Under P1, \texttt{mistral-small3.2:24b} attains the highest mean precision (0.572) while conceding little recall. Under P2, precision rises and recall falls across the board, and the precision-optimal model changes to \texttt{qwen3.6:35b}, which reaches the highest mean precision (0.634), ahead of the other leading models, whose mean precision falls between 0.59 and 0.62. The degenerate classifiers remain at the recall-only extreme under both prompts and are not considered further.

Per-repository precision varies widely even for the leading models, ranging from roughly 0.34 to 0.83 for \texttt{mistral-small3.2:24b} under P1. This reinforces that the large across-repository variance, rather than the mean ranking alone, should temper any deployment decision.

\begin{figure}[htbp]
\centering
\includegraphics[width=\columnwidth]{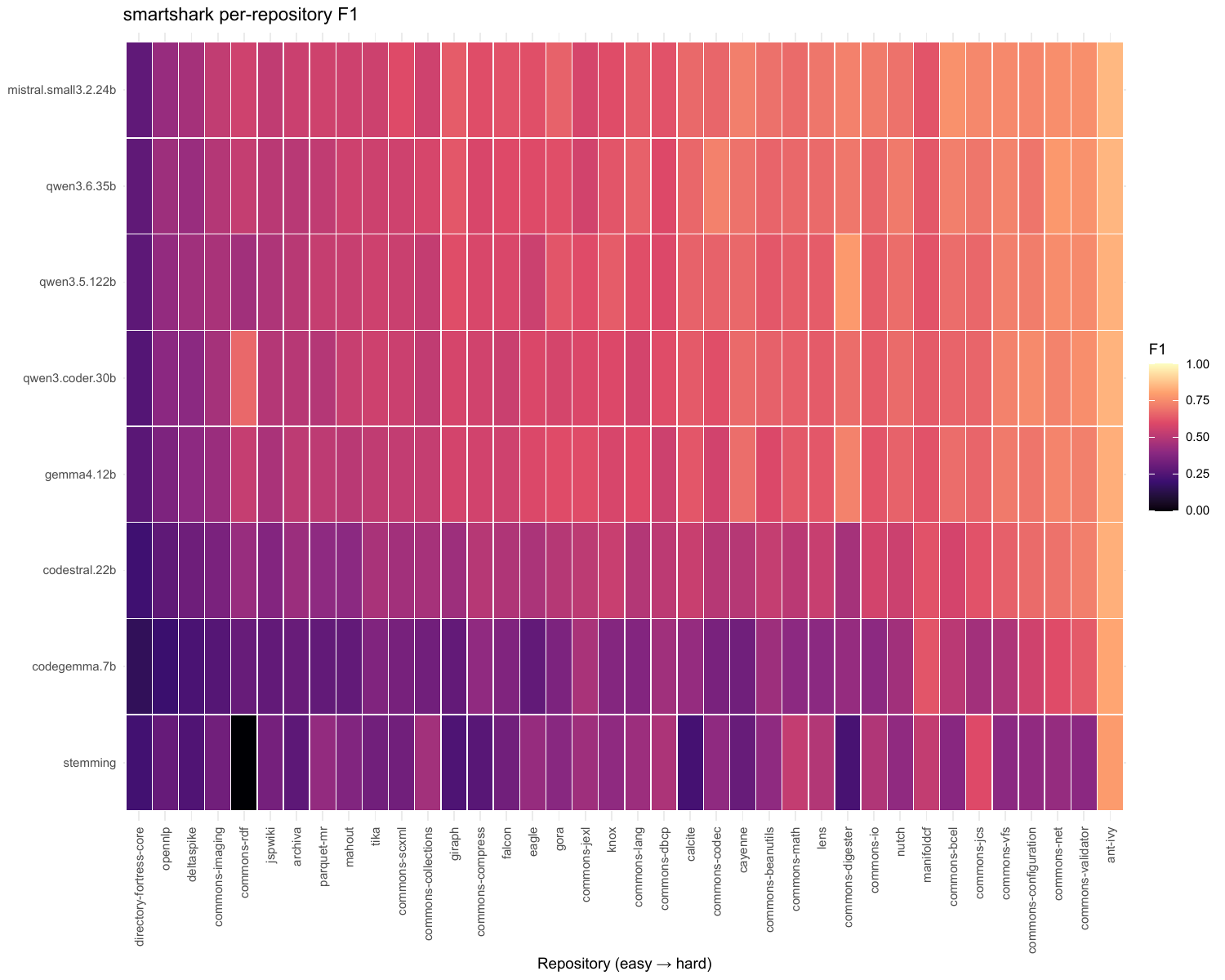}
\caption{Absolute per-repository F1 on \textsc{smartshark} under prompt
P1. Rows ordered by mean F1, columns by repository mean F1.}
\label{fig:ss-abs-p1}
\end{figure}

\begin{figure}[htbp]
\centering
\includegraphics[width=\columnwidth]{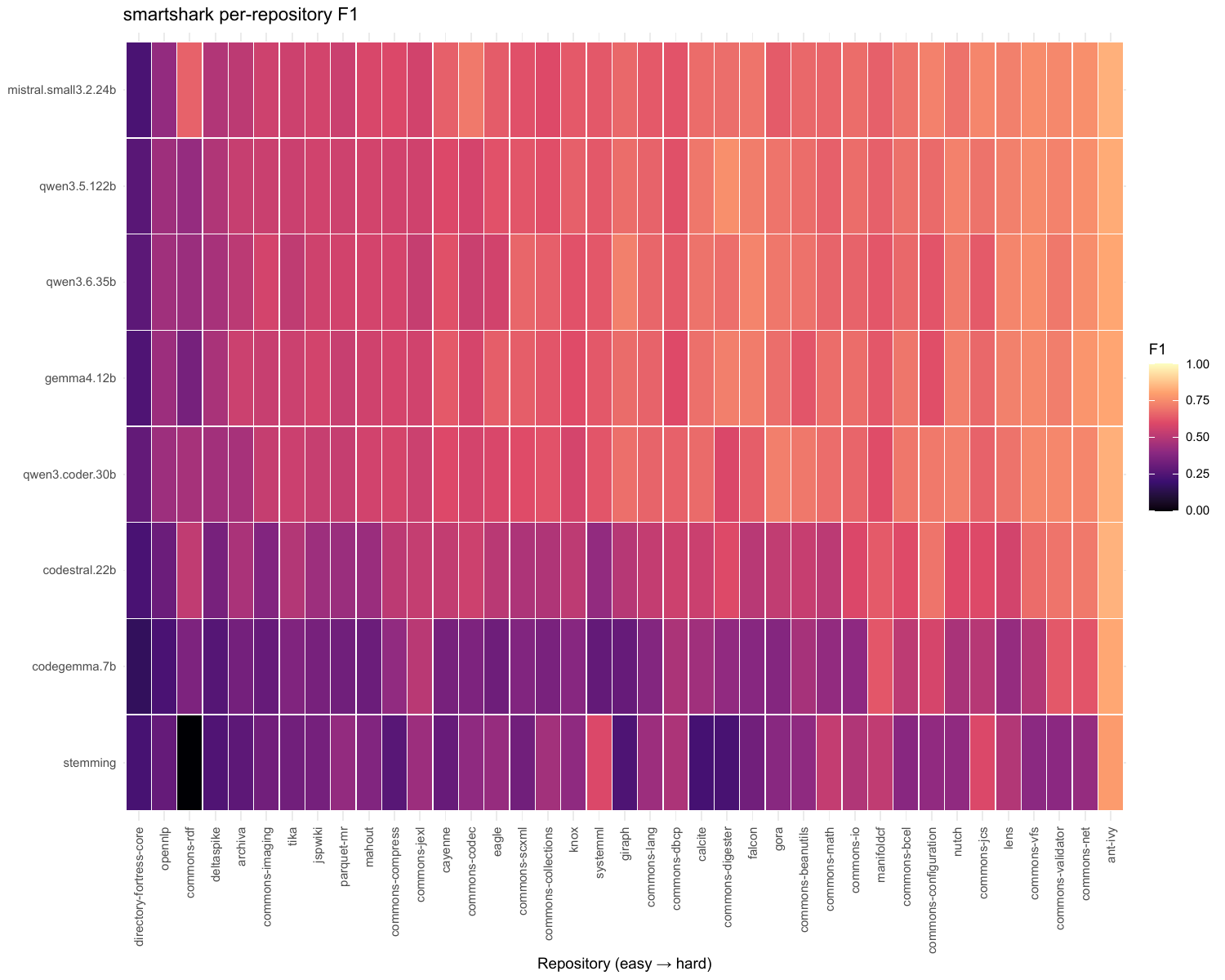}
\caption{Absolute per-repository F1 on \textsc{smartshark} under prompt
P2. The easy-to-hard repository gradient is preserved.}
\label{fig:ss-abs-p2}
\end{figure}

\begin{figure}[htbp]
\centering
\includegraphics[width=\columnwidth]{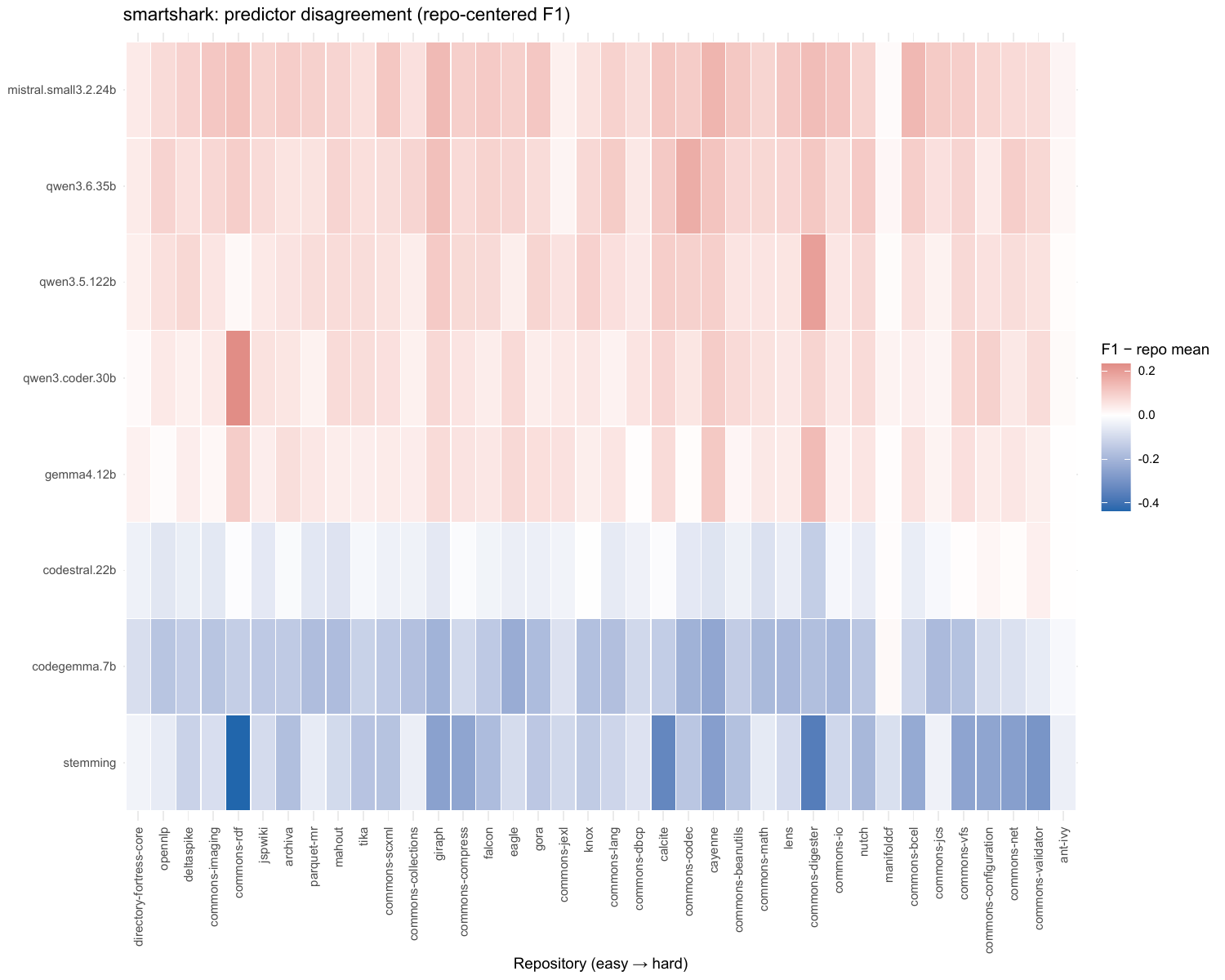}
\caption{Predictor disagreement on \textsc{smartshark} under prompt P1:
F1 with each repository's mean subtracted. The row-coherent structure
indicates a repository-independent predictor ranking.}
\label{fig:ss-rel-p1}
\end{figure}

\begin{figure}[htbp]
\centering
\includegraphics[width=\columnwidth]{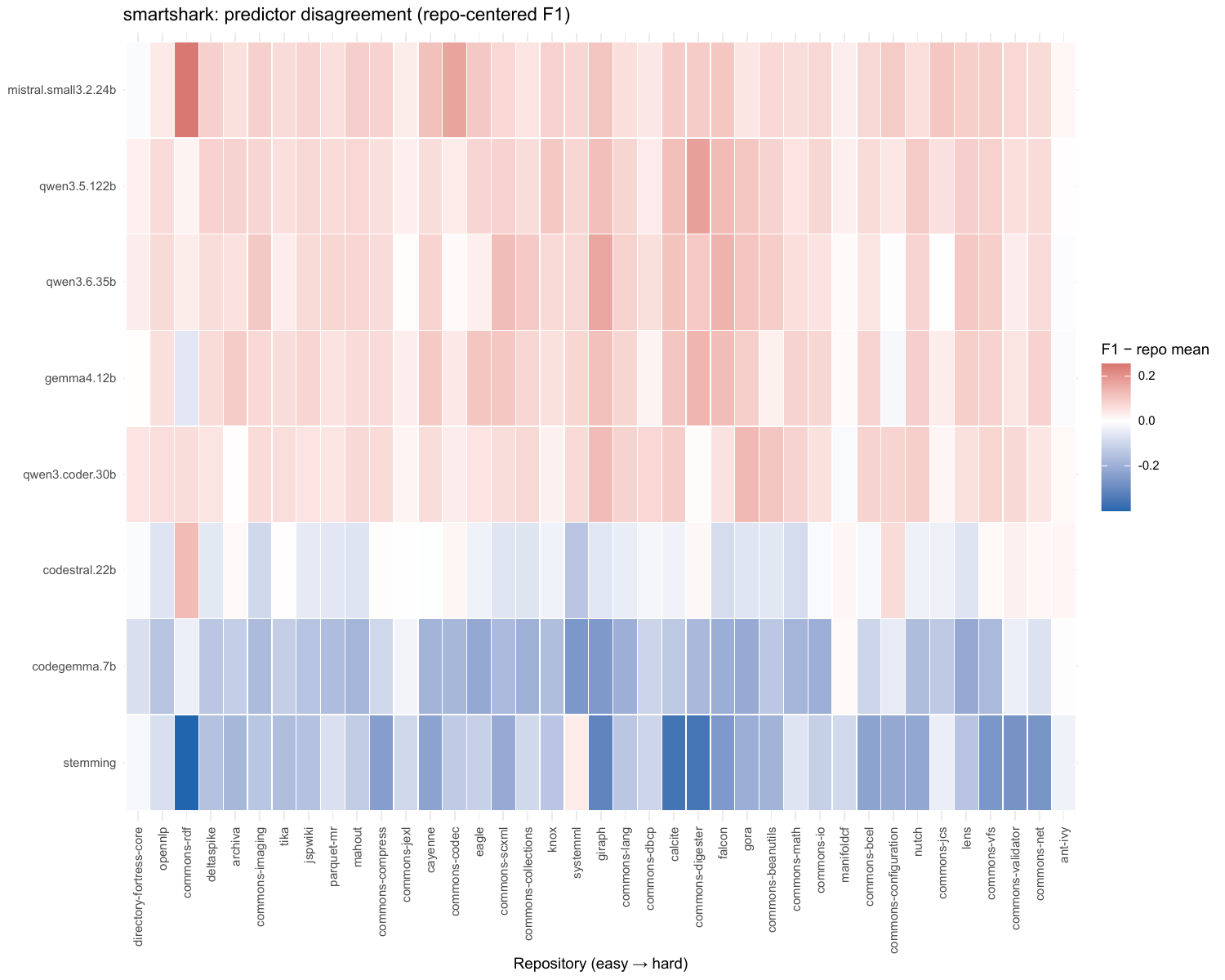}
\caption{Predictor disagreement on \textsc{smartshark} under prompt P2.
The row-coherent structure is preserved, confirming ranking stability
across prompts.}
\label{fig:ss-rel-p2}
\end{figure}

\paragraph{Spatial structure}
To distinguish repository difficulty from predictor quality, we present two complementary views of the per-repository F1-Scores on \textsc{smartshark}, for each prompt. \Cref{fig:ss-abs-p1} and \Cref{fig:ss-abs-p2} show absolute F1-Scores, while \Cref{fig:ss-rel-p1} and \Cref{fig:ss-rel-p2} show F1-Scores with each repository's across-predictor mean removed, isolating the contribution of predictor choice.

Two facts emerge cleanly and hold under both prompts. First, absolute performance is governed chiefly by the repository, as the absolute-F1-Score maps are dominated by a horizontal easy-to-hard gradient, with repositories such as \texttt{directory-fortress-core} and \texttt{opennlp} yielding low F1-Scores, while others (e.g.\ \texttt{commons-net}, \texttt{ant-ivy}) showcase high F1-Scores for every predictor. Second, and more importantly for model selection, the repository-centered maps are almost entirely \emph{row-coherent}, which means that each predictor sits above or below the repository mean by a roughly constant amount across the corpus. The leading cluster sits above the repository mean on essentially every repository, whereas \texttt{codegemma:7b} and \texttt{stemming} sit below it on essentially every repository. The predictor ranking is therefore not an averaging artifact. Instead, it holds repository by repository under both prompts, which materially strengthens the case for the leading cluster.

Departures from this pattern are rare and localised, with the most notable being \texttt{commons-rdf}, where the ordering reshuffles and where the \texttt{stemming} baseline collapses to near-zero F1-Scores (the darkest cell in the absolute maps). Such cases identify repositories whose characteristics interact with a specific predictor and are candidates for qualitative follow-up, but they do not alter the corpus-level ordering.

\subsection{Discussion} \label{sec:discussion}

\paragraph{Prompt P1}
Under P1 the classifiers separate most sharply on the recall axis. The leading models combine high recall (roughly 0.75-0.82 on \textsc{smartshark}) with moderate precision, and \texttt{mistral-small3.2:24b} attains both the best mean F1-Score and the best average Friedman rank. The post-hoc analysis shows it to be co-best with \texttt{qwen3.6:35b}, as the two models are within the critical difference, while significantly outranking \texttt{qwen3-coder:30b}, \texttt{gemma4:12b}, and the remaining models. Therefore, for P1 the recommendation is to use either of the top pair \texttt{mistral-small3.2:24b} or \texttt{qwen3.6:35b}, which is statistically separated from the rest of the field, rather than a single uncontested winner.
 
\paragraph{Prompt P2}
P2 shifts every model toward precision, as \textsc{smartshark} recall drops, by up to roughly 0.17 among the leading models, and precision rises correspondingly. This shift has a statistical consequence as the Friedman/Nemenyi analysis shows the five leading language models become mutually indistinguishable under P2 (average ranks within $0.95$, all pairwise $p>0.68$), whereas under P1 a top pair was separable from the field. P2 thus compresses the models as it moves them toward precision. It also moves the precision-optimal choice from \texttt{mistral-small3.2:24b} to \texttt{qwen3.6:35b}, which attains the highest mean precision (0.634) of any model-prompt combination. For an application that weights precision above recall, such as seeding SZZ, P2 with \texttt{qwen3.6:35b} is the precision-maximising configuration, at the cost of the lowest recall among the leading cluster. The released corpus is nonetheless labelled with \texttt{mistral-small3.2:24b} under P2 rather than with the precision-maximiser, since mistral leads on F1 under both prompts and holds the best rank, and under P2 the leading models are statistically indistinguishable, so releasing the strongest all-round model at the precision-leaning operating point costs no measurable precision while gaining robustness.
 
\paragraph{Comparison across prompts}
Four observations bear on our conclusions. We lead with the one that motivated the study.
 
First, and most importantly, \emph{\Glspl{LLM} deliver a large and statistically robust improvement over the established baseline on a problem where that baseline genuinely struggles}. On \textsc{smartshark}, the only corpus that permits a fair, precision-aware comparison, the \texttt{stemming} baseline lands near the bottom of the ranking under both prompts (average rank above $7.4$ of $8$), and every large language model in the leading group significantly outranks it under both prompts by the Nemenyi test. The margin is not marginal, as the best models exceed the baseline's F1-Score by roughly two thirds (around 0.64 against 0.40), and they do so while the baseline's own behaviour confirms the difficulty of the task, as its recall and precision are both low and its performance collapses entirely on some repositories. That keyword-based methods perform this poorly is itself evidence that bug-fix identification is a hard problem. \Glspl{LLM} clearing them so decisively, and doing so consistently repository by repository (the row-coherent structure of the per-repository analysis), is one central practical
finding of this work. Recent locally-runnable \Glspl{LLM} are not a marginal refinement of existing heuristics for this task, instead they are a qualitatively better instrument for it.

Second, this advantage is \emph{not contingent on a fortunate prompt}. The \Glspl{LLM}' dominance over the baseline holds under both P1 and P2, two prompts that differ substantially in their operating point. The usefulness of \Glspl{LLM} for bug-fix identification is therefore a property of the models, not of a single hand-tuned prompt, which is an important robustness claim given how often prompt sensitivity undermines \Gls{LLM} results.

Third, \emph{within} the leading group, the prompt determines how finely the models can be separated. Under P1 the Friedman/Nemenyi analysis isolates a top pair (\texttt{mistral-small3.2:24b}, \texttt{qwen3.6:35b}) significantly ahead of the other \Glspl{LLM}, whereas under P2 all five leaders fall within the critical difference and are statistically indistinguishable. The practical reading is reassuring rather than troubling, since model choice among the strong local \Glspl{LLM} matters only at the margin, so a practitioner can choose among a set of models suitable for deployment.
 
Fourth, \emph{the prompt shifts the operating point, and the useful point depends on the application}. The consistent move from P1 to P2 trades recall for precision, but because the two prompts differ in several respects at once we read this as a shift between two operating points rather than a single controlled lever, and we attribute it to no one change. Which point is preferable is not a property of the task but of the downstream use. Labels that seed SZZ benefit from the precision-leaning point, since a spurious fix there propagates into the bug-introducing labels, whereas a use that must recover as many true fixes as possible, and can tolerate false positives, is better served by the recall-leaning point. For this reason the corpus we release (\Cref{sec:datasets}) is labeled with P2, a choice suited to seeding SZZ rather than a claim that precision is preferable in general. That the operating point can be moved by prompt wording alone, without retraining, is itself a practical convenience.
\section{Threats to Validity and Limitations}
\label{sec:threats}

\paragraph{Construct validity}
The \textsc{smartshark} reference defines a bug fix by a validated link to a
bug-typed issue, an issue-level judgment rather than a code-level one. A commit
that repairs a defect without ever being linked to such an issue is counted a
non-fix, so a correct positive from a model is scored a false positive. Manual
validation of the issue types addresses the mistyping
of~\cite{Herzig2013Misclassification}, but not the linking incompleteness
of~\cite{Bird2009FairBalanced}. The precision we report on \textsc{smartshark}
is therefore a conservative lower bound, and the true value is at least as high.
For the same reason precision is a Java-only statement, as \textsc{smartshark}
is the only corpus with a negative class.

\paragraph{Internal validity}
P1 and P2 differ in several respects at once, so we make no claim about which
change moves the operating point and read the pair as a robustness check across
two prompts. The few-shot examples in P2 are hand-written and drawn from no
project in any corpus, so they cannot leak into the evaluation. The keyword
baseline is measured on mature, well-documented projects whose commit messages
are comparatively disciplined, which flatters it, so the gap we report is if
anything an underestimate of its shortfall on noisier histories.

\paragraph{Conclusion validity}
We base ordering claims on a Friedman test with Nemenyi post-hoc rather than on
raw mean F1-Scores, which controls for the large across-repository variance, and we run
it for both prompts. The test weights every repository equally regardless of its
commit count, consistent with the macro-averaging used throughout, so the
rankings are not a commit-volume-weighted statement. The across-repository
variance is large relative to the gaps between adjacent models, so the rankings
should be read together with it. The three bug-fix-only corpora cannot represent
a false positive and are used only to establish the benchmark-validity
inversion, not to rank models.

\paragraph{External validity}
Primary numbers are reported on the 34 \textsc{smartshark} projects with
validated issue types. The full set is shown as a contrast and shifts no
conclusion. The diff is truncated to ten files and fifty lines per file, so on
unusually large commits the model sees a partial change, though such commits are
rare and are seldom fixes. The labels we release are produced by a model and
carry its measured error, which is why we present the corpus as a silver
standard rather than as ground truth.
\section{Conclusions and Outlook}
\label{sec:conclusion}

Defect prediction inherits its labels from a prior decision about which commits
fix bugs, and the routine ways of making that decision are noisy. We asked
whether a commit can be labelled from its own content by a locally deployable
open-weight language model, reproducibly and without an issue tracker, and how
well the keyword heuristic it would replace actually performs.

On the manually validated corpus, keyword matching recovers fewer than half of
the bug-fixing commits, so the default much of the field relies on is a poor
approximation of the labels it stands in for, which answers RQ1. Locally
deployable open-weight models recover the large majority of those fixes and
outperform the baseline by roughly two thirds in F1, repository by repository
and with statistical significance, while the largest model is not the best, so
the task is within reach of a model that runs on modest local hardware, and the
two prompts we study differ in where they place the operating point rather than
in whether the models beat the baseline, which answers RQ2. Finally, the ranking
of classifiers on corpora that contain only positive examples is close to
reversed on the one corpus that also contains negatives, because an evaluation
without negatives cannot penalise a classifier that labels too much as a fix, so
such corpora cannot support a precision-aware comparison, which answers RQ3.

We make the following contributions.
\begin{itemize}
\item A measurement, across projects and three languages, of how far
keyword-based labeling sits from validated bug-fix labels, quantifying the gap
the cheap default leaves open.
\item An evaluation of seven locally deployable open-weight models against that
baseline, with a recommended configuration and the finding that classification
quality does not track parameter count across the range studied.
\item Evidence that evaluation corpora without negative examples cannot rank
bug-fix classifiers in a precision-aware way, which bears on how such methods
are validated.
\item A released, reproducible labeling pipeline and a labeled corpus of
roughly 390{,}000 commits across three languages, produced with the recommended
configuration and offered as a silver-standard resource for downstream use.
\end{itemize}

Much of defect-prediction research refines models and features on top of labels
it takes as given, yet those labels are produced by the step studied here, and
the usual shortcut for that step recovers fewer than half of the fixes it is
meant to capture. A locally run open-weight model closes most of that gap while
keeping the process reproducible and the code in-house, placing a more reliable
label at the point where the errors would otherwise originate. Because these
labels are the seed for tracing bug-introducing changes and the training signal
for the models built afterwards, a better label at the source is inherited by
everything downstream. How large that inherited improvement is, and where a
hosted frontier model would repay the reproducibility it costs, are the natural
next questions.

\section*{Acknowledgements}
The financial support by the Austrian Federal Ministry of Economy, Energy and
Tourism, the National Foundation for Research, Technology and Development and
the Christian Doppler Research Association is gratefully acknowledged.
SBA Research (SBA-K1 NGC) is a COMET Center within the COMET Competence
Centers for Excellent Technologies Programme and funded by BMIMI, BMWET, and
the federal state of Vienna. The COMET Programme is managed by FFG. This work
was funded by the Austrian Research Promotion Agency (FFG) through the BRIDGE
programme, project I-SEE, FFG project no.~933312.

\section*{Data Availability}
The labeling pipeline, prompt templates, evaluation runs, released corpus, and
statistical-analysis scripts are available at \url{https://github.com/Raubkatz/BugFixCommitLabelling}.

\bibliographystyle{IEEEtran}
\bibliography{references}

\end{document}